# Polarized X-rays from table-top free-electron-driven van der Waals crystals

Nikhil Pramanik[1*], Sunchao Huang[1*], Ruihuan Duan[2,3*], Lee Wei Wesley Wong[1], Qingwei Zhai[1], Wen Wei Lee[1], Zheng Liu[3] and Liang Jie Wong[1‡]

[1]School of Electrical and Electronic Engineering, Nanyang Technological University, 50 Nanyang Avenue, Singapore 639798, Singapore

[2]CINTRA CNRS/NTU/THALES, UMI 3288, Research Techno Plaza, Nanyang Technological University, 50 Nanyang Avenue, Singapore 637371, Singapore

[3]School of Materials Science and Engineering, Nanyang Technological University, 50 Nanyang Avenue, Singapore 639798, Singapore

‡Email: liangjie.wong@ntu.edu.sg

*These authors have contributed equally.

**Abstract:**

Polarized X-rays are essential for polarization-sensitive techniques such as X-ray absorption near-edge structure (XANES) spectroscopy, magnetic microscopy, and fluorescence imaging. These techniques are crucial in the study of molecular and material dynamics, which are in turn central to driving the global transition to clean energy and sustainability through research in photovoltaics, catalysis and optoelectronics. Currently, polarized X-rays with tunable polarization and photon energy are available only in large facilities like synchrotrons. Here, we theoretically predict and experimentally demonstrate the generation of linearly polarized X-rays from free electron-driven van der Waals crystals, without the need for additional optical stages like polarizers. We show that precise, real-time control over the polarization can be achieved by simply re-orienting the van der Waals crystal. Furthermore, we show that a high degree of polarization can be achieved across all polarization angles – from perpendicular to in-plane – with intensities exceeding $10^9$ photons $s^{-1}$ $sr^{-1}$ $mA^{-1}$. Using our findings, we present a scheme for table-top XANES, achieving performance comparable to that of synchrotron facilities. In the process, we develop a new methodology for performing XANES using high-quality polarized X-rays with larger spectral linewidths, named ORIXANES (Orientation-controlled XANES). Our findings pave the way to greater accessibility and smaller carbon footprint for advanced X-ray modalities.

## Introduction

Polarized X-rays have long been a powerful experimental probe for investigating materials and organisms in nature, offering sensitivity to anisotropic properties that unpolarized beams cannot access[1–3]. In particular, linearly polarized X-rays are essential for techniques such as X-ray absorption linear dichroism, which measure differences in absorption between orthogonal polarizations to reveal direction-specific variations in orbital occupancy and lattice geometry[4,5]. These capabilities are especially powerful for probing the electronic and structural dynamics of semiconductor nanomaterials. For instance, ultrafast X-ray absorption linear dichroism studies on semiconductor nanostructures like zinc oxide (ZnO) nanorods have provided direct observation of how photoexcited charge carriers influence electronic structure and charge transport – mechanisms that directly influence the efficiency of solar energy conversion and photocatalysis[6]. Access to such detailed, element- and orientation-specific information is essential for optimizing the efficiency and performance of devices used in solar energy generation, photovoltaics, catalysis, and many clean energy applications[7,8].

Beyond semiconductors, polarization-dependent X-ray techniques such as X-ray absorption near-edge structure (XANES)[9] spectroscopy, magnetic dichroism imaging[10], and resonant fluorescence[11] have become indispensable for studying lattice geometry, electronic states, spin textures, and molecular orientation in a broad range of materials[3,9–14]. They are also crucial for exploring magneto-optical phenomena at the atomic scale[10], as well as advanced optical and quantum-related phenomena enabled by precise polarization control[15,16].

Despite their importance, polarized X-rays with the real-time tunability in polarization and in photon energy needed for the above applications are available only at large-scale facilities like synchrotrons and X-ray free-electron lasers[17]. Such sources offer high brightness and versatile polarization control but are expensive to build, require large physical and environmental footprints, and are not widely accessible. On the other hand, table-top X-ray sources like conventional X-ray tubes are not able to produce linearly polarised X-rays except when combined with bulky external polarizing optics[16]. Due to the inherently weak interaction between X-rays and matter[18,19], these polarizers introduce significant losses, leading to low throughput and complex optical configurations. Crucially, there is no practical high-efficiency polarizer for the soft X-ray region[17], a spectral regime of particular importance because it interacts strongly with low-Z elements such as carbon, nitrogen, and oxygen, which make up the vast majority of organic and biological matter. This includes the biologically significant "water window" regime (~290-540 eV), crucial for natural contrast imaging of biological cells and soft tissues in their near-native states without staining them or requiring potentially damaging preparation[20]. Moreover, conventional X-ray tubes lack photon energy tunability, as their emission is constrained to fixed characteristic lines determined by the anode material[21]. This constraint limits the ability to perform element-specific probing with fine energy tuning at high flux – an essential requirement for near-edge absorption spectroscopic techniques[22,23]. This glaring technological gap prevents polarization-sensitive X-ray spectroscopy applications from being performed in table-top setups and on a widespread scale.

Here, we take significant strides to close this gap by showing that polarized X-rays generated from table-top free electron-driven van der Waals crystals can achieve degrees of polarization exceeding 80% across all linear polarization states, with photon fluxes above $10^9$ photons $s^{-1}$ $sr^{-1}$ $mA^{-1}$ – performance suitable for selected synchrotron applications. Specifically, we experimentally measure the output X-ray polarization by leveraging the strict relationship between X-ray intensity, polarization and crystal orientation that arises from the coherent interaction between a free electron and a periodic lattice of atoms in our scheme – this allows us to infer the X-ray polarization by measuring the X-ray intensity, circumventing the need for soft X-ray polarization detectors in our experiments. Good agreement between our experimental measurements and our theoretical predictions confirms the strict relationship between X-ray intensity, polarization and crystal orientation; and our correctness in using intensity measurements to infer the measured polarization. Subsequently, we demonstrate that precise mechanical adjustment of the van der Waals crystal's tilt angles directly controls the emitted tunable X-ray polarization, enabling continuous tuning of the polarization state from in-plane to perpendicular (out-of-plane). The flexibility of van der Waals crystals in terms of material design on the atomic scale and ability to realize designer heterostructures[24] further adds to the versatility of such a source. To illustrate the utility of such a table-top source, we implement ORIXANES (Orientation-controlled XANES) – a table-top realization of polarization-resolved X-ray absorption near-edge structure spectroscopy in which the required polarization state is deterministically engineered through crystal orientation. Using our experimentally feasible polarization quality and other parameters we simulate ORIXANES measurements on naturally occurring biomineral crystals. Our results reveal synchrotron-grade sensitivity for detecting bond orientation-specific absorption features. Our work thus paves the way for a table-top means of performing spectroscopy applications requiring polarized, tunable X-ray sources, which have so far been restricted to large X-ray facilities.

**Results and Analysis**

Van der Waals materials are layered crystals composed of atomically thin sheets held together by weak van der Waals forces (e.g., graphite). Their defining feature – dangling-bond-free surfaces – allows for the stacking of dissimilar layers without lattice-matching constraints, enabling the fabrication of van der Waals heterostructures with precisely tailored periodicities. Beyond this structural flexibility, van der Waals materials offer high in-plane thermal conductivity[25], strong electromagnetic field confinement[26], and broad tunability in both lattice spacing and elemental composition[24,27–30]. These properties make them ideal candidates for table-top tunable X-ray sources[26,31–43]. Besides, curved van der Waals structures and related nanophotonic approaches have the capability to generate focused X-ray beams[32,44,45]. Their large interlayer spacing is particularly advantageous for soft X-ray generation using table-top electron sources[33,37]. Recent studies have leveraged these characteristics of van der Waals materials to demonstrate enhanced versatility in tunable X-ray generation[34,46,47], multicolor X-ray emission[35,48], water-window X-ray generation[37], and the first experimental observation of quantum-recoil effects[36,49,50]. These advances form part of a broader range of phenomena that have recently been explored in compact and nanophotonic free-electron radiation platforms[26,51–59]. In this work, we uncover a simple yet novel technique – tuning the crystal orientation – to control X-ray polarization from free-electron interactions with van der Waals materials, enabling tunable, linearly polarized X-rays from a compact, table-top setup.

In this paper, our approach uses free-electron-driven radiation from tilted van der Waals crystals to emit linearly polarized X-rays whose polarization is determined by the van der Waals crystal orientation. By mechanically rotating the van der Waals crystal in both polar ($\theta_{\rm til}$) and azimuthal ($\phi_{\rm til}$) tilt angles with respect to the incident electron beam, we achieve precise, real-time control over the emitted X-ray polarization angle. This polarization-tunable X-ray emission arises from the coherent interaction between free electrons and a periodic array of atoms, a process that includes parametric X-ray radiation and higher order interactions such as coherent bremsstrahlung[33,34,60–65]. Parametric X-ray radiation results from the diffraction of the electron's Coulomb field by the crystal's periodic lattice, whereas coherent bremsstrahlung results from the acceleration of electrons as they undulate periodically in the crystal's electrostatic potential. Still higher-order radiation mechanisms arise from the additional diffraction of the electron's Coulomb field arising from the nonlinear parts of the electron trajectory, but tend to be negligible compared to parametric X-ray radiation and coherent bremsstrahlung, with parametric X-ray radiation typically being the dominant radiation mechanism overall in our regime of study. Henceforth in this work, we will use the term parametric X-ray radiation to refer collectively to all radiation mechanisms arising from this process of free electrons traversing a crystal lattice.

Figure 1a illustrates the polarised X-ray generation mechanism underlying our approach. As a free-electron beam traverses a layered van der Waals crystal, the coherent interaction with the periodic lattice generates X-rays that are inherently linearly polarized, with distinct polarization along different observation directions. The polar tilt angle $\theta_{\text{til}}$ inclines the crystal's [001] zone axis relative to the z-axis, and the azimuthal tilt $\phi_{\text{til}}$ rotates the crystal about the z-axis within the x-y plane. Each unique $(\theta_{\text{til}}, \phi_{\text{til}})$ pair therefore defines a specific crystal orientation. A strict, deterministic link among crystal orientation, emitted intensity, and polarization enables a new polarization measurement methodology, in which angular-resolved intensity measurements over $\theta_{\text{til}}$ and $\phi_{\text{til}}$ are used to quantitatively infer the polarization state, which we explore in the following sections.

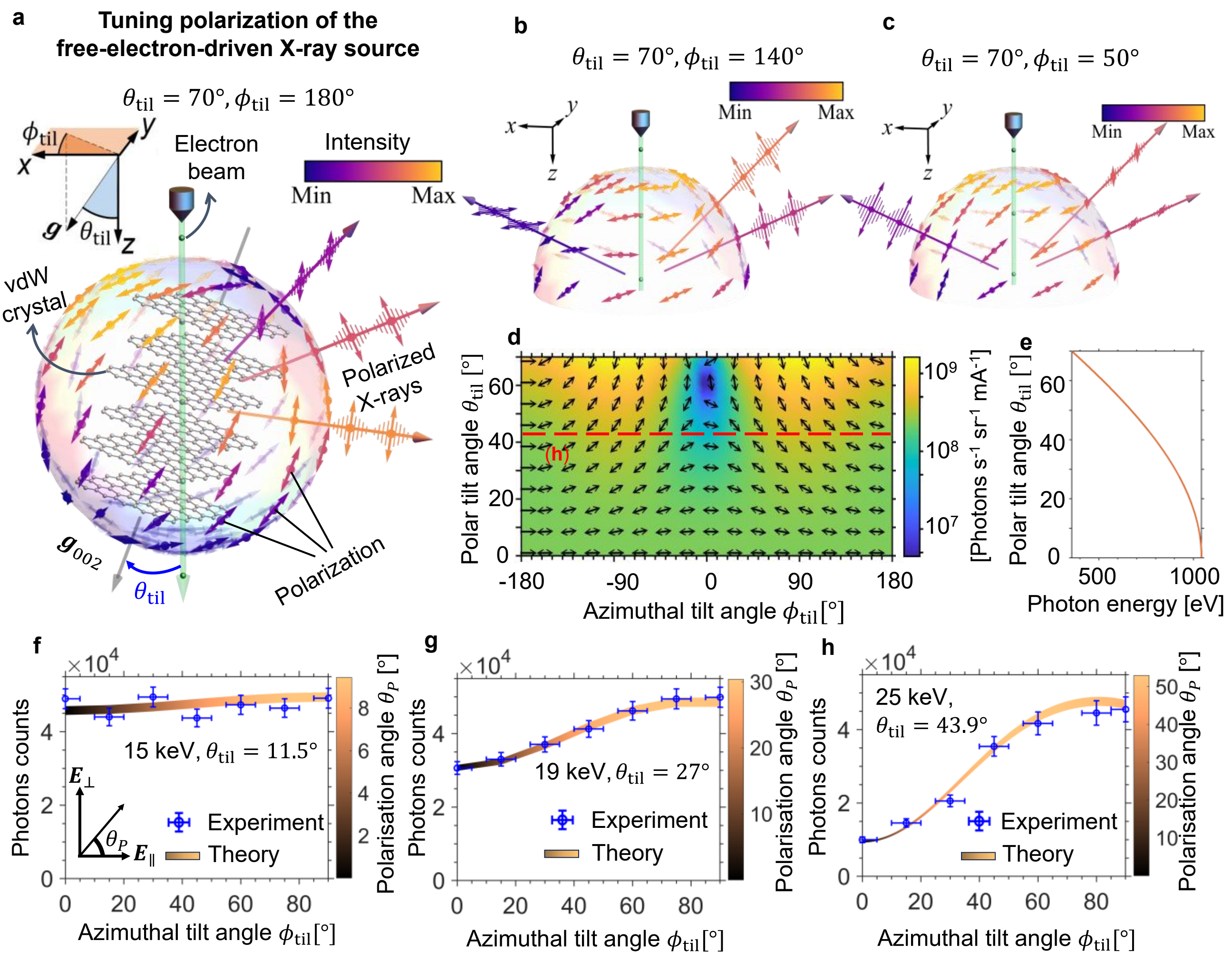

**Fig. 1. Linearly Polarized X-rays from free-electron-driven van der Waals crystals. a**, Schematic illustration of X-ray emission generated by free electrons passing through a van der Waals crystal. The coherent interaction between the electrons and the crystal's periodic atomic lattice produces linearly polarized X-rays. The polarization direction varies with observation angle, indicated by double-headed arrows, while the X-ray intensity is represented by the color scale (arbitrary units). **b**, **c**, Variation of the azimuthal tilt angle ($\phi_{\mathrm{til}}$) of the van der Waals crystal modulates the polarization direction of the emitted X-rays at fixed observation angles. The three arrows in each panel illustrate how the polarization – denoted by the electric field oscillation direction – changes with crystal rotation. **d**, Simulated X-ray intensity colormap as a function of the polar ($\theta_{\mathrm{til}}$) and azimuthal ($\phi_{\mathrm{til}}$) tilt angles. The color scale indicates the emitted X-ray intensity, while overlaid arrows denote the corresponding polarization directions. The colormap reveals deterministic and continuous control of the polarization direction through crystal orientation ($\theta_{\mathrm{til}}$, $\phi_{\mathrm{til}}$): horizontal and vertical double-headed arrows correspond to pure in-plane and perpendicular polarization, respectively, while inclined arrows represent mixed polarization states. Simulation is performed for a 100 nm graphite crystal under 20 keV incident electron energy. **e**, X-ray energy as a function of polar tilt angle, corresponding to panel d. The emission energy remains independent of $\phi_{\mathrm{til}}$, demonstrating that the polarization can be tuned without affecting the X-ray energy. **f-h**, Experimental validation of the intrinsic relationship between X-ray intensity, polarization angle, and crystal orientation. X-ray photon counts is experimentally measured as the crystal is azimuthally rotated (varying $\phi_{\mathrm{til}}$) for the polar tilt angles $\theta_{\mathrm{til}} = 11.5°$, 27°, and 43.9° in f, g and h, respectively. Theoretical predictions (shaded, color-coded curves) show the expected intensity variation are in excellent agreement with the experimentally observed intensity modulations. The strong agreement between experimental measurements and theoretical predictions substantiates the fundamental principle that polarization, intensity, and crystal orientation are strictly interlinked in the coherent interaction between free electrons and periodic lattice of atoms. The shaded curve is obtained via theory simulations using Eq. (2), where the color of the curve denote the polarization angle. The incident electron energy is 15 keV, 19 keV, and 25 keV for f, g and h, respectively. The observation angle of the X-ray detector is $\theta_{\mathrm{obs}} = 119°$, which corresponds to an elevation angle of 29°.

In the parametric X-ray radiation mechanism, the peak photon energy $E_{\mathrm{pk}}$ of the emitted radiation is determined by

$$E_{\mathrm{pk}} = \hbar \frac{\boldsymbol{v} \cdot \boldsymbol{g}_{\mathrm{rot}}(\theta_{\mathrm{til}}, \phi_{\mathrm{til}})}{1 - (\boldsymbol{v} \cdot \hat{\boldsymbol{k}} / \mathrm{c})}, \quad \boldsymbol{g}_{\mathrm{rot}}(\theta_{\mathrm{til}}, \phi_{\mathrm{til}}) = \hat{R}(\theta_{\mathrm{til}}, \phi_{\mathrm{til}})\, \boldsymbol{g}, \tag{1}$$

where $\hbar$ is the reduced Planck constant, $\boldsymbol{v}$ is the velocity of the incident electron, $\hat{\boldsymbol{k}}$ is the unit vector along the observation direction, c is the speed of light in free space, $\theta_{\mathrm{til}}$ and $\phi_{\mathrm{til}}$ are the angles controlling the crystal's physical orientation, and $\boldsymbol{g}_{\mathrm{rot}}(\theta_{\mathrm{til}}, \phi_{\mathrm{til}})$ is the rotated reciprocal lattice vector resulting from the change in crystal orientation. The rotation is applied via the composite matrix

$\hat{R}(\theta_{\text{til}}, \phi_{\text{til}}) = R_z(\phi_{\text{til}}) \cdot R_y(\theta_{\text{til}})$, which acts on $\boldsymbol{g}$ the reciprocal lattice vector in the unrotated frame (i.e., when $\theta_{\text{til}} = \phi_{\text{til}} = 0°$). Here, $R_z(\phi_{\text{til}})$ is the unitary rotation matrix corresponding to the azimuthal rotation of the crystal about the z-axis by an angle $\phi_{\text{til}}$, effectively rotating the crystal within the x–y plane. $R_y(\theta_{\text{til}})$ is the unitary rotation matrix that describes the polar tilt of the crystal about the y-axis by an angle $\theta_{\text{til}}$. As the crystal orientation changes, so does the rotated reciprocal lattice vector $\boldsymbol{g}_{\text{rot}}(\theta_{\text{til}}, \phi_{\text{til}})$, which directly governs the peak photon energy via the projection $\boldsymbol{v} \cdot \boldsymbol{g}_{\text{rot}}(\theta_{\text{til}}, \phi_{\text{til}})$ in the numerator of Eq. (1), while the denominator $1 - (\boldsymbol{v} \cdot \hat{\boldsymbol{k}} / \text{c})$ captures the angular dependence due to the observation direction.

Our study reveals that in addition to energy tuning, this geometric control of crystal orientation governs the polarization state of the emitted X-rays. For a fixed crystal orientation, the polarization angle varies as a function of observation angle due to the directional dependence of the radiation fields generated by the electron-crystal interaction, as illustrated in Fig. 1a. However, by rotating the crystal – particularly through azimuthal rotation $\phi_{\text{til}}$ – we unlock a second degree of freedom that allows control over the polarization direction at a fixed observation point. Figures 1b and 1c demonstrate this: for a constant polar tilt $\theta_{\text{til}} = 70°$, varying the azimuthal rotation angle from $\phi_{\text{til}} = 140°$ to $50°$ results in a clear rotation of the X-ray polarization, as indicated by the direction of the field oscillation vectors. This ability to tune polarization by independently varying either the observation direction or the crystal orientation provides a flexible and deterministic method for accessing arbitrary linear polarization states, a capability not demonstrated in previous free-electron-driven X-ray sources.

To analyse the full parameter space, we simulate X-ray emission from a 100 nm-thick graphite crystal driven by a 25 keV electron beam. We fix the observation direction at $\theta_{\text{obs}} = 119°$, $\phi_{\text{obs}} = 180°$ (experimental conditions), and map the emitted X-ray flux as a function of the tilt angles $\theta_{\text{til}}$ and $\phi_{\text{til}}$. The results are presented in Fig. 1d, where the color represents the photon flux and the overlaid double-headed arrows denote the polarization direction of the emitted X-rays. Horizontal double-headed arrows denote parallel (in-plane) polarization, vertical double-arrows denote perpendicular (out-of-plane) polarization, and inclined double-headed arrows denote polarizations as mixtures of the former two polarizations. The polarization direction varies smoothly and predictably with crystal tilt angles, confirming that crystal orientation provides a reliable handle for tuning the polarization vector.

In addition to polarization control, our simulations reveal a distinct intensity modulation of the emitted X-ray intensity as a function of crystal orientation. As shown in Fig. 1d, the X-ray intensity reaches a maximum when the azimuthal tilt $\phi_{\text{til}}$ approaches $\pm 90°$, for a fixed polar tilt $\theta_{\text{til}}$. This intensity variation arises purely from the geometric configuration of the crystal, without any change in electron energy or observation direction. The enhancement reflects the deterministic dependence of both intensity and polarization on the relative orientation between the electron velocity vector $\boldsymbol{v}$, the rotated reciprocal lattice vector $\boldsymbol{g}_{\text{rot}}(\theta_{\text{til}}, \phi_{\text{til}})$, and the emission direction $\hat{\boldsymbol{k}}$.

To quantitatively explain the observed angular dependence of intensity and polarization, we examine the parametric X-ray radiation mechanism in greater detail. For accurate simulations in our study, we employ a theoretical framework, which combines first-principles electromagnetism with Monte Carlo simulations of electron scattering[5]. The total emitted X-ray intensity is obtained by summing the radiation contributions from all electron trajectory segments as they interact with the periodic crystal potential. The differential photon yield is given by:

$$\frac{d^2N}{d\omega d\Omega} = \frac{\alpha\, \omega}{N_e 4\pi^2 c^2} \sum_{j=1}^{N_e} \sum_{i=1}^{N_{Coll}^{j}} \left| \int_0^{t_L} \boldsymbol{v}_i(t) \cdot \boldsymbol{E}_{\boldsymbol{ks}}(\boldsymbol{r}_i(t), \omega)\, e^{-i\omega t}\, dt \right|^2, \quad (2)$$

where $N$ is the number of emitted photons, $\omega$ is the angular frequency, $\Omega$ is the solid angle, $N_e$ is the number of incident electrons, $\alpha$ is the fine-structure constant, $N_{\text{Coll}}^{j}$ denotes the number of scattering events along the $j^{\text{th}}$ electron trajectory, $t_L$ is the interaction time for the $i^{\text{th}}$ electron trajectory segment, $\boldsymbol{v}_i$ and $\boldsymbol{r}_i$ is the velocity and position of the $i^{\text{th}}$ electron trajectory segment. The electromagnetic crystal eigenmode $\boldsymbol{E}_{\boldsymbol{k}s}$, corresponding to wavevector $\boldsymbol{k}$ and polarization vector $\hat{\mathbf{e}}_s$, is given by

$$\boldsymbol{E}_{\boldsymbol{k}s}(\boldsymbol{r},\omega) = e^{i\boldsymbol{k}\cdot\boldsymbol{r}}\,\hat{\mathbf{e}}_s + \sum_{\boldsymbol{g}} -\frac{\chi_g}{\left(k_g^2 - \frac{\omega^2}{c^2}\right)}\left[\boldsymbol{k}_{\boldsymbol{g}}\,(\boldsymbol{g}_{\mathrm{rot}}(\theta_{\mathrm{til}},\phi_{\mathrm{til}})\cdot\hat{\mathbf{e}}_s) - \frac{\omega^2}{c^2}\hat{\mathbf{e}}_s\right]e^{i(\boldsymbol{k}+\boldsymbol{g})\cdot\boldsymbol{r}}, \tag{3}$$

where, $\chi_g$ is the electric susceptibility, and $\boldsymbol{k}_{\boldsymbol{g}} = \boldsymbol{k} + \boldsymbol{g}_{\mathrm{rot}}$. The parallel and perpendicular polarization components are defined as $\widehat{\mathbf{e}_{\parallel}} = \widehat{\boldsymbol{k}} \times \hat{\mathbf{z}}$ and $\widehat{\mathbf{e}_{\perp}} = \widehat{\boldsymbol{k}} \times \widehat{\mathbf{e}_{\parallel}}$, where $\hat{\mathbf{z}}$ is the unit vector along z-axis direction and $\widehat{\boldsymbol{k}}$ is the unit vector along the observation direction. The polarization angle $\theta_p$ is defined as the acute angle between the X-ray electric-field vector and $\widehat{\mathbf{e}_{\parallel}}$, the in-plane polarization axis. The angular dependence of X-ray intensity and polarization angle revealed by Eqs. (2) and (3) can be understood by examining the role of the crystal eigenmode $\boldsymbol{E}_{\boldsymbol{k}s}$. In particular, the coupling term $\boldsymbol{g}_{\mathrm{rot}} \cdot \hat{\mathbf{e}}_s$ governs the polarization-dependent modulation of the field. When the emitted wavevector $\boldsymbol{k}$ becomes parallel (or nearly parallel) to the rotated reciprocal lattice vector $\boldsymbol{g}_{\mathrm{rot}}$, this dot product vanishes, suppressing the contribution from the term $\boldsymbol{k}_{\boldsymbol{g}}(\boldsymbol{g}_{\mathrm{rot}} \cdot \hat{\mathbf{e}}_s)$ in Eq. (3). As a result, the X-ray intensity drops sharply – a phenomenon we observe experimentally as a dip in flux near certain orientations (i.e., when $\phi_{\mathrm{til}} = 0°$). By introducing an azimuthal rotation $\phi_{\mathrm{til}}$, we break this alignment between $\boldsymbol{k}$ and $\boldsymbol{g}_{\mathrm{rot}}$, thereby restoring the coupling and reactivating the suppressed term. This geometric reconfiguration leads to a strong boost in intensity, accounting for the observed flux enhancement in Fig. 1d when $\phi_{\mathrm{til}}$ approaches $\pm 90°$. Crucially, this deterministic relationship between crystal orientation, emission intensity, and polarization forms the basis of the polarization measurement methodology introduced in this work. By exploiting the predictable variation of intensity with tilt angles, the polarization state of the emitted X-rays can be quantitatively inferred from angular-resolved intensity measurements.

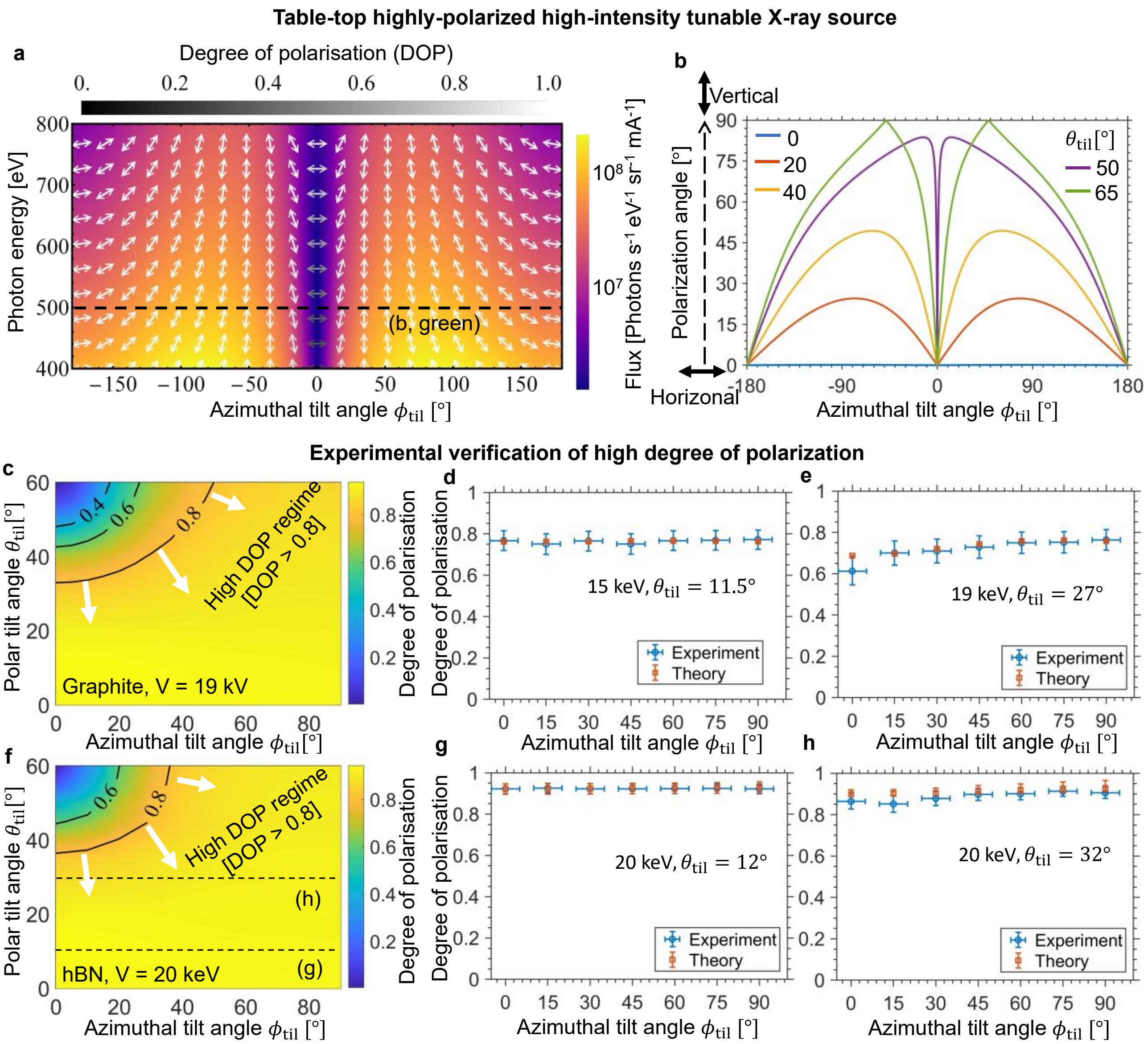

**Fig. 2. High-flux emission, full polarization tunability and high degree of polarization (> 80%) in table-top free electron-driven van der Waals crystals. a,** Degree of polarization (DOP) colormap as a function of peak photon energy and azimuthal tilt angle $\phi_{til}$. High degree of polarization is observed when the crystal is azimuthally rotated away from $\phi_{til} = 0°$, with X-rays collected at an observation angle $\theta_{obs} = 119°$. This indicates that azimuthal rotation is essential for achieving high polarization, in contrast to conventional approaches where the azimuthal angle is typically fixed at 0°. Here, the polar tilt angle $\theta_{til}$ is fixed at 65°, where full polarization tunability is accessible. **b** shows the smooth tunability of polarization angle of the emitted X-ray with varying azimuthal angle for a 500 eV photon energy. **c-e,** Experimental verification of the degree of polarization for a 77.5±1 nm thick graphite thin film. The experimental data, denoted by blue, uses the peak X-ray spectrum value as measured by the EDS detector, where the detector resolution is convoluted with the intrinsic parametric X-ray radiation peak. The bremsstrahlung background for the degree of polarization calculation is obtained from numerical simulations based on Monte Carlo electron scattering – obtained from NIST DTSA II software[67,68]. The electron energy used for b and c is 15 keV and 19 keV respectively and the polar tilt angle is 11.5° and 27°, respectively. **f-h** show the experimental verification of degree of polarization for 85±9 nm thick hBN thin film.

To validate our polarization measurement methodology, we experimentally measure the X-ray flux as a function of azimuthal tilt $\phi_{\text{til}}$ for three different polar tilt angles: $\theta_{\text{til}} = 11.5°$, $27°$, and $44°$ (Figs. 1f-1h). The X-ray measurements are carried out in a field emission scanning electron microscope (FESEM) that provided a source of free electrons. The output X-ray spectra are measured by an energy dispersive spectrometer (EDS) detector positioned at an elevation angle of $29°$ ($\theta_{\text{obs}} = 119°$). The experimental data are shown in Figs. 1f–1h as solid dots with error bars, overlaid with shaded theoretical curves obtained from simulations based on Eq. (2), where the color of the shaded region encodes the polarization angle of the emitted X-rays at each azimuthal configuration. This representation simultaneously captures the flux variation and polarization evolution, enabling a direct mapping between crystal orientation, polarization state, and emission intensity. This interdependence, governed by Eqs. (2) and (3), is clearly validated in Figs. 1f–1h, confirming that azimuthal rotation of the crystal allows deterministic and continuous tuning of both the polarization angle and intensity at fixed photon energy.

Our table-top polarized X-ray source achieves a degree of polarization (DOP) exceeding 80% across the full range of linear polarization states, while maintaining high photon flux suitable for tackling selected synchrotron applications. The degree of polarization is defined as $\text{DOP} = I_{\text{pol}}/(I_{\text{pol}} + I_{\text{unpol}})$, where $I_{\text{pol}}$ is the flux of the polarized parametric X-ray radiation and $I_{\text{unpol}}$ is the flux of the unpolarized incoherent bremsstrahlung background. Figure 2a presents a simulated colormap of degree of polarization as a function of photon energy and azimuthal tilt angle, where the overlaid arrows indicate the polarization and the color denotes the value of degrees of polarization. Horizontal and vertical double-headed arrows represent pure in-plane and perpendicular polarizations, respectively. Across a wide range of tilt angles, the degree of polarization remains above 0.8 (> 80%), demonstrating that polarization purity is preserved throughout the tunable range. Notably, high degrees of polarization are achieved only when the crystal is azimuthally rotated away from the conventional configuration ($\phi_{\text{til}}$ = 0°). In contrast to traditional approaches, where the crystal is typically aligned without azimuthal rotation, such configurations correspond to reduced polarization purity in our system. Our results therefore reveal that azimuthal rotation is not merely a tuning parameter, but a necessary operating condition for accessing high polarization purity in free-electron-driven van der Waals X-ray sources. In addition to maintaining high degree of polarization, our source enables continuous and smooth tuning of the polarization angle at each photon energy. As shown in Fig. 2b, this angle can be tuned continuously from vertical to horizontal polarization. The range of accessible polarization angles depends on the polar tilt angle $\theta_{\mathrm{til}}$ – larger polar tilt angles yield broader tunability. In our simulation, full polarization coverage is achieved when polar tilt angle $\theta_{\mathrm{til}}$ exceeds 62°, providing access to all linear polarization states.

Figure 2c shows the simulated colormap of degree of polarization of our X-rays as a function of polar and azimuthal tilt angles, showing high degree of polarization across a broad range of crystal orientations. To experimentally confirm the predicted high degree of polarization, we perform X-ray measurement experiments from graphite and hexagonal boron nitride (hBN) samples under different tilt angles and electron energies, as shown in Figs. 2d and 2e. The experimental results closely match theoretical predictions, confirming the generation of highly polarized X-rays across the tunable range. Figures 2f and 2g extend the validation to hBN crystals. These results confirm that our source provides robust, tunable, and highly polarized X-rays across different materials and operating conditions.

Figure 3 introduces ORIXANES (Orientation-controlled XANES) — a methodology for performing polarization-resolved near-edge spectroscopy using a broader-linewidth table-top X-ray source. Unlike conventional XANES, which relies on highly monochromatic synchrotron radiation, ORIXANES reconstructs high-resolution spectra from a sequence of polarization-resolved broad-linewidth inputs with tunable peak photon energy. We evaluate this approach using biomineral crystals, demonstrating through simulations that ORIXANES reproduces key near-edge spectral features comparable to synchrotron-based measurements[66], while operating with substantially reduced infrastructure and power requirements. In our simulation, a 100-nm-thick calcite biomineral crystal is irradiated with polarized X-rays derived from our polarized X-ray source (Fig. 3a), and the transmitted spectra are recorded using a soft X-ray emission spectrometer. In our front-to-end (source to detector) simulations, we account for our source's broader-linewidth spectrum, electron scattering in the target crystal, material self-absorption, detector noise during X-ray collection, and the attenuation losses due to the X-ray optics inside the spectrometer. Owing to the high degree of polarization attainable in our platform, the resulting XANES spectra closely approach those obtained using synchrotron radiation.

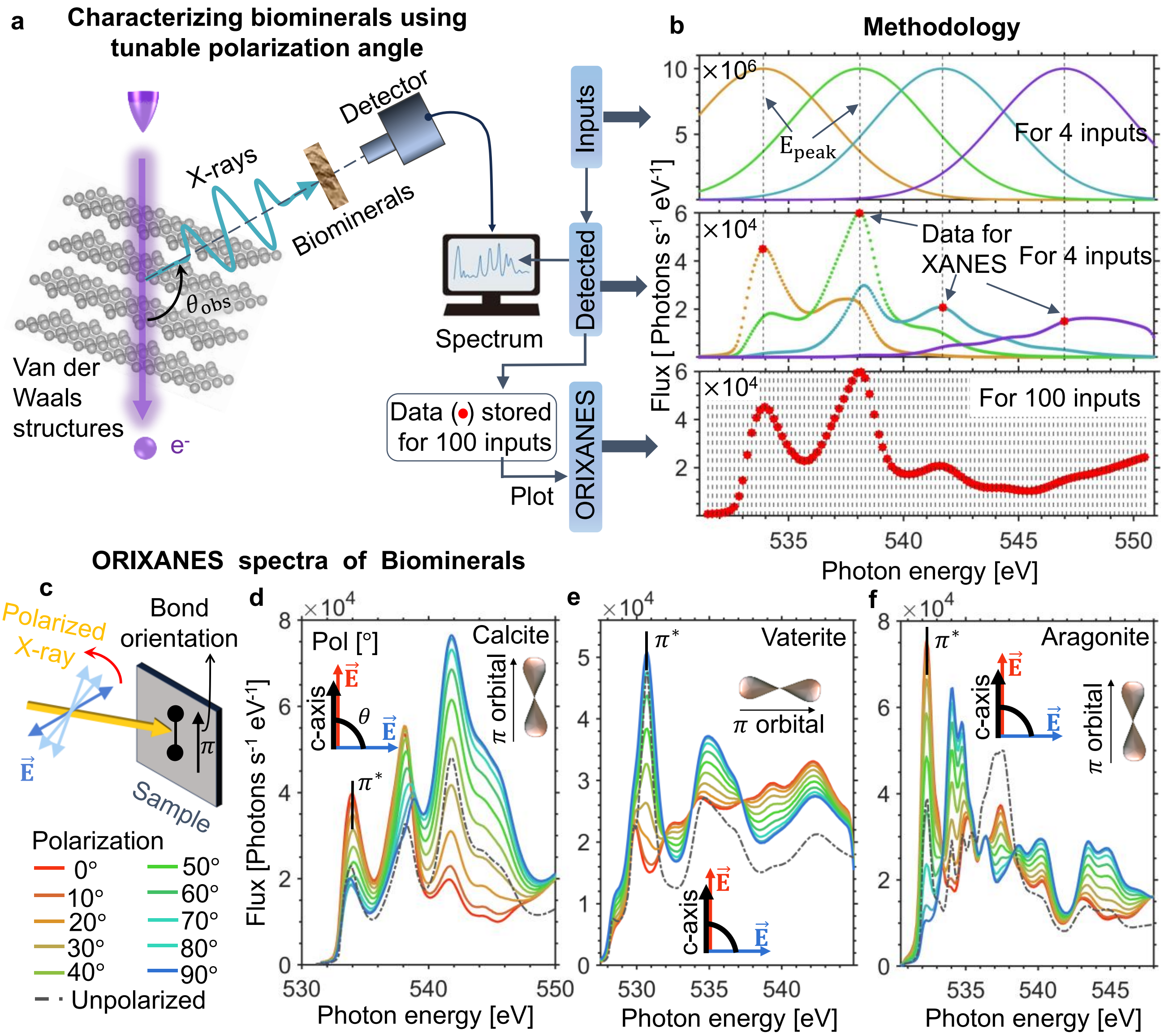

**Fig. 3. Table-top X-ray absorption near-edge structure (XANES) spectroscopy of biominerals, enabled by free electron-driven van der Waals crystals. a** shows the schematic diagram of our proposed ORIXANES experimental setup**.** Polarized X-rays emitted from our source pass through a 100 nm-thick biomineral crystal. The absorption of these X-rays within the crystal depends on their polarization, influenced by the bond orientations within the crystal – a phenomenon known as X-ray linear dichroism. X-ray linear dichroism describes the difference in absorption of X-rays that are linearly polarized parallel versus perpendicular to a specific orientation axis within the material. This effect provides insights into the orientation and electronic properties of the material's bonds. Maximum absorption occurs when the X-ray polarization aligns with the bond orientation, while it is minimized when perpendicular. The resulting ORIXANES spectrum reveals detailed information about the bond orientations and types present in the sample. **b** outlines our methodology for acquiring ORIXANES spectra using a broader-linewidth spectrum. We have utilized a 90–102 keV tunable electron beam incident on a graphite crystal to generate polarized X-rays between 530 eV and 550 eV. The peak photon energy along any observation direction is determined by Eq. (1), which allows tuning through three parameters: the incident electron-beam energy, the crystal tilt angle, and the observation angle. For the four cases shown in b, we used electron energies of 92.1 keV, 94.4 keV, 96.7 keV, and 99.6 keV to produce peak photon energies of 534 eV, 538 eV, 542 eV, and 547 eV, respectively, all at a crystal tilt angle of 70°. The crystal is then azimuthally rotated to sample the full range of polarization states. The emitted spectrum from our source illuminates the biomineral, and the transmitted X-rays are collected using a soft X-ray emission spectrometer. This spectrometer captures a convoluted spectrum, shown in the "Detected" section. We record the detected intensity data at the peak photon energy of the input spectrum, repeating this process for 100 inputs to generate 100 ORIXANES data points (shown by red dots). These measurements are assembled across the sequence of inputs to reconstruct the ORIXANES spectrum, as shown in the "ORIXANES" section. **c** shows the schematic diagram of polarized X-rays falling on a biomineral crystal and the color of the horizontal bar denote the X-ray polarization angle. **d, e, and f** display simulated ORIXANES spectra for three calcium carbonate biominerals: calcite (d), vaterite (e), and aragonite (f). In these simulations, the crystal c-axis is vertical, and the orientation of the linear polarization vector (**E**) varies from vertical ($\theta = 0°$, denoted by red, where θ is the angle between the c-axis and **E**) to horizontal ($\theta = 90°$, denoted by blue) in 10° increments. The vertical black lines indicate the experimental $\pi^*$ peak positions. For calcite and aragonite, the $\pi^*$ peak is at its maximum at $\theta = 0°$, while for vaterite, the $\pi^*$ peak is at its maximum at $\theta = 90°$. This behavior is consistent with the crystal structures of these minerals. The $\pi^*$ peak corresponds to the O 1s $\rightarrow \pi^*$ C=O transition. This transition involves the excitation of an oxygen 1s electron to the $\pi^*$ anti-bonding orbital of the carbonate group ($CO_3^{2-}$). The $\pi^*$ orbitals are perpendicular to the plane of the carbonate group.

To generate the polarized X-rays used for ORIXANES, we have utilized an incident electron beam tunable from 90 keV to 102 keV, directed onto a 100 nm thick graphite crystal, to produce tunable, polarized X-ray emission in the 530-550 eV range – required for the application studied in Figure 3. The peak photon energy along a given observation direction is determined by Eq. (1), enabling three independent photon energy tuning mechanisms: the electron-beam energy, the crystal tilt angle, and the observation angle. For the four inputs shown in Fig. 3b, we used electron energies of 92.1 keV, 94.4 keV, 96.7 keV, and 99.6 keV to obtain peak photon energies of 534 eV, 538 eV, 542 eV, and 547 eV, respectively, where the crystal tilt angle $\theta_{\text{til}}$ was fixed at 70°. The crystal was subsequently azimuthally rotated to generate the full range of polarization states (in-plane to out-of-plane). With an incident electron beam current of about 2 mA – 7 mA, a flux of $10^8$ photons $s^{-1}$ $eV^{-1}$ $sr^{-1}$ was generated for different photon energies at each polarization state.

Due to the broader spectral linewidth associated with our radiation mechanism compared to synchrotron sources, direct measurement of high-resolution near-edge spectra is not possible using conventional approaches. This fundamental difference necessitates a new spectroscopy framework. To address this, we introduce ORIXANES (Orientation-controlled XANES), a dedicated measurement and reconstruction methodology for XANES tailored to broader-linewidth X-ray sources whose polarization can be controlled through crystal orientation, as outlined in Fig. 3b.

In ORIXANES, each input spectrum is defined by a chosen crystal orientation and peak photon energy, such that the incident polarization state and peak photon energy are jointly controlled. Rather than measuring absorption at a single monochromatic energy, the transmitted intensity sampled at the peak photon energy of each input is assembled over many incrementally tuned inputs to reconstruct the polarization-dependent near-edge spectrum. The ORIXANES procedure is as follows. The polarized emission from the van der Waals source illuminates the biomineral sample, and the transmitted X-rays are captured by a soft X-ray emission spectrometer, resulting in a convoluted spectrum (labeled "Detected"). To build the ORIXANES spectrum, we measure the transmitted intensity at the peak photon energy of each input spectrum and repeat this process across 100 inputs, each with incrementally shifted peak photon energies. The resulting set of peak flux data points (red dots) forms the reconstructed ORIXANES spectrum, as shown in Fig. 3b.

Figure 3d shows the reconstructed ORIXANES spectrum from biomineral calcite crystals using our polarized X-ray source, revealing bond alignment and bond types. Maximum absorption occurs when the X-ray polarization aligns with specific bond directions, while perpendicular alignment minimizes absorption. For example, the $\pi^*$ transition peak (O 1s → $\pi^*$ C=O), reaches its highest flux at $\theta = 0°$ (red curve), confirming the orientation of the $\pi$ orbitals perpendicular to the carbonate plane. These polarization-dependent absorption features extracted via ORIXANES provide direct insight into the anisotropic electronic structure of calcite and its relation to molecular bonding.

Figures 3e and 3f present reconstructed ORIXANES spectra for other naturally occurring forms of calcium carbonate biominerals, vaterite and aragonite respectively, demonstrating distinct polarization-dependent absorption profiles for each biomineral. The variation in the $\pi^*$ peak position and intensity among these structures highlights the differences in their molecular arrangements and bonding environments. In particular, while calcite and aragonite exhibit maximum $\pi^*$ peak absorption at $\theta = 0°$, vaterite shows the opposite trend, with its highest absorption at $\theta = 90°$ (blue curve). This observation is consistent with the structural orientations of the carbonate groups within each crystal, reaffirming the sensitivity of our X-ray source to bond anisotropy in biominerals. For comparison, the grey dotted curve represents the spectrum obtained using an unpolarized X-ray source. In this case, all polarization states contribute equally, resulting in a blended spectrum where directional information is lost. As a result, the ability to distinguish bond orientation is suppressed.

These results establish the feasibility of performing XANES outside of synchrotron facilities using a compact, table-top source. The continuous energy tunability and precise polarization control of our platform, together with the ORIXANES methodology, enable detailed investigations of crystallographic orientation, molecular bonding, and electronic anisotropy across a broad range of natural and synthetic materials within a standard laboratory environment.

## Discussion and conclusion

Our source is highly complementary to existing table-top X-ray generation methods[69]. Laser-based sources such as high-harmonic generation (HHG) provide ultrafast X-ray pulses, but they rely on complex, high-power laser systems and their emission is typically limited to the soft X-ray regime. In contrast, our source relies on low-energy electron beams to access a broad photon energy range without the need for high-power laser setups. Conventional electron-impact sources, such as X-ray tubes, emit unpolarized, fixed-energy characteristic X-rays determined by the anode material and require external polarizers to produce polarized beams. These polarizers introduce significant losses due to the weak interaction between X-rays and matter. Our source eliminates the need for external polarizing optics and intrinsically generates polarized X-rays with real-time tunability of both X-ray energy and polarization. By simply adjusting the azimuthal and polar tilt angles of the van der Waals crystal, we achieve precise polarization control, high flux ($10^9$ photons $s^{-1}$ $sr^{-1}$ $mA^{-1}$), and a degree of polarization exceeding 80% for all polarization states.

Our work bridges the gap between large-scale synchrotron facilities and laboratory-based X-ray sources. In particular, ORIXANES establishes a practical framework for performing polarization-resolved XANES using compact, broader-linewidth and tunable X-ray sources by combining deterministic polarization control and spectral reconstruction, retaining sensitivity to bond-orientation-dependent absorption features and achieving spectral contrast comparable to synchrotron-based measurements in targeted studies. A particularly important application of our system is in pre-screening samples for synchrotron experiments. While synchrotrons offer high flux and highly monochromatic beams, they suffer from limited beamline availability and long user waiting times. Our tabletop source, available at own laboratory, allows researchers to rapidly evaluate and prioritize samples or experimental conditions before committing to scarce synchrotron time.

An additional strength of our X-ray generation platform lies in the intrinsic design flexibility of van der Waals materials. Unlike conventional crystals that require lattice matching, van der Waals materials can be stacked layer-by-layer to form heterostructures with tailored interlayer spacing and composition, which directly influences the X-ray emission spectrum. Recent experiments have shown that stacking dissimilar van der Waals layers, such as $MoS_2$ and graphite, can yield multicolor X-ray spectra, with individual peaks attributed to each material[35]. This spectral control at the material-level complements our polarization tuning via crystal orientation, collectively facilitating compact X-ray sources with user-defined spectral profiles and polarization states.

The combined polarization control, energy tunability, and compactness of our platform open avenues for novel applications. For instance, coupling the source with pulsed electron beams could enable time-resolved polarized X-ray studies of dynamic processes in chemical and biological sciences on a table-top setup. Our method of X-ray generation also lays the foundation for fully integrated, chip-scale X-ray systems. On the optics side, chip-scale X-ray waveguides and zone-plate-based diffractive architectures provide complementary routes to integrated beam delivery[70–74]. Ongoing progress in miniaturized electron sources[75–82] may soon make on-chip polarized X-ray emitters a practical reality. The integration of high brightness, tabletop scalability, energy tunability, and precise control over arbitrary X-ray polarization makes this technology a cornerstone for next-generation compact X-ray systems.

In conclusion, we have theoretically and experimentally demonstrated a table-top mechanism for generating tunable, linearly polarized X-rays from free-electron interactions with van der Waals crystals. We showed that crystal orientation provides deterministic control over the emitted polarization, and experimentally validated the strict relationship between crystal orientation, X-ray intensity, and polarization through the agreement between measured and predicted intensity modulations. Our source achieves degrees of polarization exceeding 80% across the full range of linear polarization states, with photon fluxes in excess of $10^9$ photons $s^{-1}$ $sr^{-1}$ $mA^{-1}$. Building on these capabilities, we introduced ORIXANES, a methodology for performing polarization-resolved near-edge spectroscopy using broader-linewidth X-ray sources. Unlike conventional approaches that rely on highly monochromatic radiation, ORIXANES enables reconstruction of polarization-dependent spectral features from inputs with finite spectral linewidth, where both the polarization state and peak photon energy are deterministically defined. We therefore establish a compact platform for polarization-resolved X-ray spectroscopy and imaging, bringing capabilities previously restricted to large-scale synchrotron facilities into a more accessible table-top format for selected applications.

## Methods

X-ray measurements: The X-ray measurements were performed in a FESEM JEOL7800. The emitted X-ray spectra were measured using a silicon drift energy dispersive X-ray spectroscopy (EDS) detector in the FESEM. The EDS detector was calibrated to measure the X-ray photon energies with an accuracy within ± 1 eV by measuring the $K_\alpha$ peaks of C, N, B, O, F & Si. The detector area is 170 $mm^2$ where the sample to detector distance is kept at 92 mm. The background radiation of the measured spectra was subtracted using NIST DTSA-II software[67,68].

Material synthesis and preparation: The graphite nanoflakes are exfoliated mechanically onto silicon substrates (with 285 nm $SiO_2$ film) and transferred to Au grids with the aid of the wet-transfer method. The Au grid is held by a copper TEM grid holder.

## Acknowledgements

We thank C. Boothroyd, A. Lim and Y. Y. Tay, for helpful discussions. This project is supported by the Ministry of Education, Singapore, under its AcRF Tier 2 programme (award no. MOE-T2EP50222-0012) and Tier 1 programme (award no. RG139/23). We acknowledge the Facility for Analysis, Characterisation, Testing and Simulation, Nanyang Technological University, Singapore, for use of their electron microscopy/X-ray facilities.

## Author contributions

N.P. led the project. N.P. and S.H. designed and performed the electron microscopy experiments. R.D. and Z.L. prepared the samples. N. P. developed the theory and performed the simulations. N.P., S.H.,

L.W.W.W., Z.Q. and W.W.L. analysed the experimental and simulation results. N.P. and L.J.W. wrote the paper, with inputs from all other authors. L.J.W., N.P. and S.H. conceived the idea. L.J.W. supervised the project.

### Competing interests

The authors declare no competing interests.

## References

1. Van Bokhoven, J. A. & Lamberti, C. *X-Ray Absorption and X-Ray Emission Spectroscopy, 2 Volume Set: Theory and Applications*. vol. 1 (John Wiley & Sons, 2016).
2. Stöhr, J. *NEXAFS Spectroscopy*. (Springer Science & Business Media, 2013).
3. Salditt, T., Egner, A. & Luke, D. R., Eds., *Nanoscale Photonic Imaging*. vol. 134 (Springer International Publishing, Cham, 2020).
4. Huang, D. J., Wu, W. B., Guo, G. Y., Chen, C. T., Lin, H.-J., Hou, T. Y., Chang, C. F., Fujimori, A., Kimura, T., Huang, H. B., Tanaka, A. & Jo, T. Nature of orbital ordering in La0.5Sr1.5MnO4 studied by soft x-ray linear dichroism. in *14th International Conference on Vacuum-Ultraviolet Radiation Physics. Program and Abstracts* 234–234 (2004).
5. Hageraats, S., Thoury, M., Stanescu, S. & Keune, K. Soft X-ray nanospectroscopy for quantification of X-ray linear dichroism on powders. *J Synchrotron Rad* **28**, 1090–1099 (2021).
6. Rossi, T. C., Dykstra, C. P., Haddock, T. N., Wallick, R., Burke, J. H., Gentle, C. M., Doumy, G., March, A. M. & van der Veen, R. M. Charge Carrier Screening in Photoexcited Epitaxial Semiconductor Nanorods Revealed by Transient X-ray Absorption Linear Dichroism. *Nano Lett.* **21**, 9534–9542 (2021).
7. Groot, F. de & Kotani, A. *Core Level Spectroscopy of Solids*. (CRC Press, Boca Raton, 2008).
8. Electronic and Magnetic Interactions in Solids. in *Magnetism: From Fundamentals to Nanoscale Dynamics* (eds Stöhr, J. & Siegmann, H. C.) 221–310 (Springer, Berlin, Heidelberg, 2006).
9. Guttmann, P., Bittencourt, C., Rehbein, S., Umek, P., Ke, X., Van Tendeloo, G., Ewels, C. P. & Schneider, G. Nanoscale spectroscopy with polarized X-rays by NEXAFS-TXM. *Nature Photon* **6**, 25–29 (2012).
10. Fischer, P. Frontiers in imaging magnetism with polarized x-rays. *Front. Phys.* **2**, (2015).
11. Chi, Z., Du, Y., Huang, W. & Tang, C. Linearly polarized X-ray fluorescence computed tomography based on a Thomson scattering light source: a Monte Carlo study. *J Synchrotron Rad* **27**, 737–745 (2020).
12. Chauvin, M., Florén, H.-G., Friis, M., Jackson, M., Kamae, T., Kataoka, J., Kawano, T., Kiss, M., Mikhalev, V., Mizuno, T., Ohashi, N., Stana, T., Tajima, H., Takahashi, H., Uchida, N. & Pearce, M. Shedding new light on the Crab with polarized X-rays. *Sci Rep* **7**, 7816 (2017).
13. Spezzani, C., Allaria, E., Coreno, M., Diviacco, B., Ferrari, E., Geloni, G., Karantzoulis, E., Mahieu, B., Vento, M. & De Ninno, G. Coherent Light with Tunable Polarization from Single-Pass Free-Electron Lasers. *Phys. Rev. Lett.* **107**, 084801 (2011).

14. Ma, Y., Hua, J., Liu, D., He, Y., Zhang, T., Chen, J., Yang, F., Ning, X., Zhang, H., Du, Y. & Lu, W. Compact Polarized X-Ray Source Based on All-Optical Inverse Compton Scattering. *Phys. Rev. Applied* **19**, 014073 (2023).
15. Döpp, A., Mahieu, B., Lifschitz, A., Thaury, C., Doche, A., Guillaume, E., Grittani, G., Lundh, O., Hansson, M., Gautier, J., Kozlova, M., Goddet, J. P., Rousseau, P., Tafzi, A., Malka, V., Rousse, A., Corde, S. & Ta Phuoc, K. Stable femtosecond X-rays with tunable polarization from a laser-driven accelerator. *Light Sci Appl* **6**, e17086–e17086 (2017).
16. Schulze, K. S., Grabiger, B., Loetzsch, R., Marx-Glowna, B., Schmitt, A. T., Garcia, A. L., Hippler, W., Huang, L., Karbstein, F., Konôpková, Z., Schlenvoigt, H.-P., Schwinkendorf, J.-P., Strohm, C., Toncian, T., Uschmann, I., Wille, H.-C., Zastrau, U., Röhlsberger, R., Stöhlker, T., Cowan, T. E. & Paulus, G. G. Towards perfectly linearly polarized x-rays. *Phys. Rev. Research* **4**, 013220 (2022).
17. Miyawaki, J., Yamamoto, S., Hirata, Y., Horio, M., Harada, Y. & Matsuda, I. Fast and versatile polarization control of X-ray by segmented cross undulator at SPring-8. *AAPPS Bull.* **31**, 25 (2021).
18. Hart, M. & Rodrigues, A. R. D. Tuneable polarizers for X-rays and neutrons. *Philosophical Magazine B* **40**, 149–157 (1979).
19. Hart, M., Siddons, D. P., Amemiya, Y. & Stojanoff, V. Tunable x-ray polarimeters for synchrotron radiation sources. *Rev. Sci. Instrum.* **62**, 2540–2544 (1991).
20. Kördel, M., Dehlinger, A., Seim, C., Vogt, U., Fogelqvist, E., Sellberg, J. A., Stiel, H. & Hertz, H. M. Laboratory water-window x-ray microscopy. *Optica* **7**, 658 (2020).
21. Als-Nielsen, J. & McMorrow, D. *Elements of Modern X-Ray Physics*. (John Wiley & Sons, 2011).
22. Henderson, G. S., De Groot, F. M. & Moulton, B. J. X-ray absorption near-edge structure (XANES) spectroscopy. *Reviews in Mineralogy and Geochemistry* **78**, 75–138 (2014).
23. Newton, M. A., Zimmermann, P. & Van Bokhoven, J. A. X-Ray Absorption Spectroscopy (XAS): XANES and EXAFS. in *Springer Handbook of Advanced Catalyst Characterization* (eds Wachs, I. E. & Bañares, M. A.) 565–600 (Springer International Publishing, Cham, 2023).
24. Geim, A. K. & Grigorieva, I. V. Van der Waals heterostructures. *Nature* **499**, 419–425 (2013).
25. Jiang, P., Qian, X., Gu, X. & Yang, R. Probing Anisotropic Thermal Conductivity of Transition Metal Dichalcogenides $MX_2$ (M = Mo, W and X = S, Se) using Time-Domain Thermoreflectance. *Advanced Materials* **29**, 1701068 (2017).
26. Wong, L. J., Kaminer, I., Ilic, O., Joannopoulos, J. D. & Soljačić, M. Towards graphene plasmon-based free-electron infrared to X-ray sources. *Nature Photon* **10**, 46–52 (2016).
27. Jin, G., Lee, C.-S., Okello, O. F. N., Lee, S.-H., Park, M. Y., Cha, S., Seo, S.-Y., Moon, G., Min, S. Y., Yang, D.-H., Han, C., Ahn, H., Lee, J., Choi, H., Kim, J., Choi, S.-Y. & Jo, M.-H.

Heteroepitaxial van der Waals semiconductor superlattices. *Nat Nanotechnol* **16**, 1092–1098 (2021).

28. Li, J., Yang, X., Liu, Y., Huang, B., Wu, R., Zhang, Z., Zhao, B., Ma, H., Dang, W., Wei, Z., Wang, K., Lin, Z., Yan, X., Sun, M., Li, B., Pan, X., Luo, J., Zhang, G., Liu, Y., Huang, Y., Duan, X. & Duan, X. General synthesis of two-dimensional van der Waals heterostructure arrays. *Nature* **579**, 368–374 (2020).
29. Liang, S.-J., Cheng, B., Cui, X. & Miao, F. Van der Waals Heterostructures for High-Performance Device Applications: Challenges and Opportunities. *Advanced Materials* **32**, 1903800 (2020).
30. Zhang, Z., Lin, P., Liao, Q., Kang, Z., Si, H. & Zhang, Y. Graphene-Based Mixed-Dimensional van der Waals Heterostructures for Advanced Optoelectronics. *Adv. Mater.* **31**, 1806411 (2019).
31. Balanov, A., Gorlach, A. & Kaminer, I. Temporal and spatial design of x-ray pulses based on free-electron–crystal interaction. *APL Photonics* **6**, 070803 (2021).
32. Shi, X., Kurman, Y., Shentcis, M., Wong, L. J., García De Abajo, F. J. & Kaminer, I. Free-electron interactions with van der Waals heterostructures: a source of focused X-ray radiation. *Light Sci Appl* **12**, 148 (2023).
33. Shentcis, M., Budniak, A. K., Shi, X., Dahan, R., Kurman, Y., Kalina, M., Herzig Sheinfux, H., Blei, M., Svendsen, M. K., Amouyal, Y., Tongay, S., Thygesen, K. S., Koppens, F. H. L., Lifshitz, E., García de Abajo, F. J., Wong, L. J. & Kaminer, I. Tunable free-electron X-ray radiation from van der Waals materials. *Nat. Photonics* **14**, 686–692 (2020).
34. Huang, S., Duan, R., Pramanik, N., Boothroyd, C., Liu, Z. & Wong, L. J. Enhanced Versatility of Table-Top X-Rays from Van der Waals Structures. *Advanced Science* **9**, 2105401 (2022).
35. Huang, S., Duan, R., Pramanik, N., Go, M., Boothroyd, C., Liu, Z. & Wong, L. J. Multicolor x-rays from free electron–driven van der Waals heterostructures. *Sci. Adv.* **9**, eadj8584 (2023).
36. Huang, S., Duan, R., Pramanik, N., Herrin, J. S., Boothroyd, C., Liu, Z. & Wong, L. J. Quantum recoil in free-electron interactions with atomic lattices. *Nat. Photon.* **17**, 224–230 (2023).
37. Pramanik, N., Huang, S., Duan, R., Zhai, Q., Go, M., Boothroyd, C., Liu, Z. & Wong, L. J. Fundamental scaling laws of water-window X-rays from free-electron-driven van der Waals structures. *Nat. Photon.* **18**, 1203–1211 (2024).
38. Zhai, Q., Pramanik, N., Duan, R., Huang, S., Liu, Z. & Wong, L. J. Enhanced tunable X-rays from bulk crystals driven by table-top free electron energies. *Nat Commun* **16**, 11218 (2025).
39. Roques-Carmes, C., Kooi, S. E., Yang, Y., Rivera, N., Keathley, P. D., Joannopoulos, J. D., Johnson, S. G., Kaminer, I., Berggren, K. K. & Soljačić, M. Free-electron–light interactions in nanophotonics. *Appl. Phys. Rev.* **10**, 011303 (2023).

40. Wong, L. J., Huang, S., Pramanik, N., Zhai, Q. & Go, M. Method and System for Generating X-Ray Emission in the Water Window Regime. International patent application WO2024123241. (2024).

41. Wong, L. J., Huang, S. & Boothroyd, C. Apparatus and Methods for Generating Tunable X-Rays Via the Interaction of Free Electrons with Periodic Structures. International patent application WO2023027644. (2023).

42. Balanov, A., Gorlach, A. & Kaminer, I. Breaking the barriers of electron-driven x-ray radiation in crystals. *APL Photonics* **9**, 076101 (2024).

43. Zhai, Q., Pramanik, N., Duan, R., Huang, S., Liu, Z. & Wong, L. J. Enhanced X-Ray Generation from Bulk van der Waals Materials. in *IEEE Photonics Conference (IPC), pp.* 1–2 (2025).

44. Shi, X., Shentcis, M., Kurman, Y., Wong, L. J., García De Abajo, F. J. & Kaminer, I. Free-electron-driven X-ray caustics from strained van der Waals materials. *Optica* **10**, 292 (2023).

45. Karnieli, A., Roitman, D., Liebtrau, M., Tsesses, S., Van Nielen, N., Kaminer, I., Arie, A. & Polman, A. Cylindrical Metalens for Generation and Focusing of Free-Electron Radiation. *Nano Lett.* **22**, 5641–5650 (2022).

46. Pramanik, N., Huang, S., Qingwei, Z., Go, M., Duan, R., Boothroyd, C., Liu, Z. & Wong, L. J. Fundamental Scaling Laws for Tunable X-ray Smith-Purcell Radiation in Van der Waals Structures. in *Conference on Lasers and Electro-Optics, paper* FTh5B.6 (2023).

47. Pramanik, N., Huang, S., Duan, R., Boothroyd, C., Liu, Z. & Wong, L. J. Tunable Table-Top X-Rays from Tilted Van Der Waals Crystals. in *Conference on Lasers and Electro-Optics, paper* FTh4A.3 (2022).

48. Huang, S., Duan, R., Pramanik, N., Go, M., Boothroyd, C., Liu, Z. & Wong, L. J. Tunable multicolor X-rays from free-electron-driven van der Waals heterostructures. in *Conference on Lasers and Electro-Optics, paper* FW4M-3 (2023).

49. Huang, S., Duan, R., Pramanik, N., Herrin, J. S., Boothroyd, C., Liu, Z. & Wong, L. J. Quantum recoil in free electron-driven spontaneous emission from van der Waals crystals. in *Conference on Lasers and Electro-Optics, paper* FF4C.4 (2022).

50. Tsesses, S., Bartal, G. & Kaminer, I. Light generation via quantum interaction of electrons with periodic nanostructures. *Phys. Rev. A* **95**, 013832 (2017).

51. Yang, Y., Roques-Carmes, C., Kooi, S. E., Tang, H., Beroz, J., Mazur, E., Kaminer, I., Joannopoulos, J. D. & Soljačić, M. Photonic flatband resonances for free-electron radiation. *Nature* **613**, 42–47 (2023).

52. Rosolen, G., Wong, L. J., Rivera, N., Maes, B., Soljačić, M. & Kaminer, I. Metasurface-based multi-harmonic free-electron light source. *Light Sci Appl* **7**, 64 (2018).

53. Christopher, J., Taleb, M., Maity, A., Hentschel, M., Giessen, H. & Talebi, N. Electron-driven photon sources for correlative electron-photon spectroscopy with electron microscopes. *Nanophotonics* **9**, 4381–4406 (2020).
54. Yang, Y., Massuda, A., Roques-Carmes, C., Kooi, S. E., Christensen, T., Johnson, S. G., Joannopoulos, J. D., Miller, O. D., Kaminer, I. & Soljačić, M. Maximal Spontaneous Photon Emission and Energy Loss from Free Electrons. *Nature Phys* **14**, 894–899 (2018).
55. Roques-Carmes, C., Kooi, S. E., Yang, Y., Massuda, A., Keathley, P. D., Zaidi, A., Yang, Y., Joannopoulos, J. D., Berggren, K. K., Kaminer, I. & Soljačić, M. Towards integrated tunable all-silicon free-electron light sources. *Nat Commun* **10**, 3176 (2019).
56. Shi, X., Lee, W. W., Karnieli, A., Lohse, L. M., Gorlach, A., Wong, L. W. W., Salditt, T., Fan, S., Kaminer, I. & Wong, L. J. Quantum nanophotonics with energetic particles: X-rays and free electrons. *Progress in Quantum Electronics* 100577 (2025).
57. Wong, L. W. W., Shi, X., Karnieli, A., Lim, J., Kumar, S., Carbajo, S., Kaminer, I. & Wong, L. J. Free-electron crystals for enhanced X-ray radiation. *Light Sci Appl* **13**, 29 (2024).
58. Wong, L. J., Rivera, N., Murdia, C., Christensen, T., Joannopoulos, J. D., Soljačić, M. & Kaminer, I. Control of quantum electrodynamical processes by shaping electron wavepackets. *Nature communications* **12**, 1700 (2021).
59. Talebi, N. Interaction of electron beams with optical nanostructures and metamaterials: from coherent photon sources towards shaping the wave function. *Journal of Optics* **19**, 103001 (2017).
60. Feranchuk, I. D. & Ivashin, A. V. Theoretical investigation of the parametric X-ray features. *J. Phys. France* **46**, 1981–1986 (1985).
61. Feranchuk, I. D., Ulyanenkov, A., Harada, J. & Spence, J. C. H. Parametric x-ray radiation and coherent bremsstrahlung from nonrelativistic electrons in crystals. *Phys. Rev. E* **62**, 4225–4234 (2000).
62. Baryshevsky, V. G. & Feranchuk, I. D. Parametric X-rays from ultrarelativistic electrons in a crystal: theory and possibilities of practical utilization. *Journal de Physique* **44**, 913–922 (1983).
63. Baryshevsky, V. G., Batrakov, K. G., Feranchuk, I. D., Grubich, A. O., Gurinovich, A. A., Lobko, A. S., Rouba, A. A., Safronov, P. F., Stolyarsky, V. I., Tarnopolsky, B. A. & Ulyanenkov, A. P. Coherent bremsstrahlung and parametric X-ray radiation from nonrelativistic electrons in a crystal. *Tech. Phys. Lett.* **32**, 392–395 (2006).
64. Baryshevsky, V. G., Feranchuk, I. D. & Ulyanenkov, A. P. *Parametric X-Ray Radiation in Crystals: Theory, Experiments, and Applications*. (Springer, 2005).

65. Afanasenko, V. P., Baryshevsky, V. G., Zuevsky, R. F., Lobko, A. S., Moskatelnikov, A. A., Nurushev, S. B., Panov, V. V., Potsilujko, V. P., Rykalin, V. V., Skorokhod, S. V. & Shvarkov, D. S. Detection of proton parametric x-ray radiation in silicon. *Physics Letters A* **170**, 315–318 (1992).
66. DeVol, R. T., Metzler, R. A., Kabalah-Amitai, L., Pokroy, B., Politi, Y., Gal, A., Addadi, L., Weiner, S., Fernandez-Martinez, A., Demichelis, R., Gale, J. D., Ihli, J., Meldrum, F. C., Blonsky, A. Z., Killian, C. E., Salling, C. B., Young, A. T., Marcus, M. A., Scholl, A., Doran, A., Jenkins, C., Bechtel, H. A. & Gilbert, P. U. P. A. Oxygen Spectroscopy and Polarization-Dependent Imaging Contrast (PIC)-Mapping of Calcium Carbonate Minerals and Biominerals. *J. Phys. Chem. B* **118**, 8449–8457 (2014).
67. Ritchie, N. W. M., Newbury, D. E. & Davis, J. M. EDS measurements of X-ray intensity at WDS precision and accuracy using a silicon drift detector. *Microscopy and Microanalysis* **18**, 892–904 (2012).
68. Newbury, D. E. & Ritchie, N. W. M. Performing elemental microanalysis with high accuracy and high precision by scanning electron microscopy/silicon drift detector energy-dispersive X-ray spectrometry (SEM/SDD-EDS). *J Mater Sci* **50**, 493–518 (2015).
69. Stein, G. J., Keathley, P. D., Krogen, P., Liang, H., Siqueira, J. P., Chang, C.-L., Lai, C.-J., Hong, K.-H., Laurent, G. M. & Kärtner, F. X. Water-window soft x-ray high-harmonic generation up to the nitrogen K-edge driven by a kHz, 2.1 μm OPCPA source. *J. Phys. B: At. Mol. Opt. Phys.* **49**, 155601 (2016).
70. Salditt, T., Hoffmann, S., Vassholz, M., Haber, J., Osterhoff, M. & Hilhorst, J. X-Ray Optics on a Chip: Guiding X Rays in Curved Channels. *Phys. Rev. Lett.* **115**, 203902 (2015).
71. Smith, H. I. & Schattenburg, M. L. X-ray lithography from 500 to 30 nm: X-ray nanolithography. *IBM J. Res. & Dev.* **37**, 319–329 (1993).
72. Smith, H. I. A proposal for maskless, zone-plate-array nanolithography. *Journal of Vacuum Science & Technology B: Microelectronics and Nanometer Structures Processing, Measurement, and Phenomena* **14**, 4318–4322 (1996).
73. Smith, H. I. Low cost nanolithography with nanoaccuracy. *Physica E: Low-dimensional Systems and Nanostructures* **11**, 104–109 (2001).
74. Smith, H. I., Schattenburg, M. L., Hector, S. D., Ferrera, J., Moon, E. E., Yang, I. Y. & Burkhardt, M. X-ray nanolithography: Extension to the limits of the lithographic process. *Microelectronic Engineering* **32**, 143–158 (1996).
75. Floreani, F., Koops, H. W. & Elsäßer, W. Concept of a miniaturised free-electron laser with field emission source. *Nuclear Instruments and Methods in Physics Research Section A: Accelerators, Spectrometers, Detectors and Associated Equipment* **483**, 488–492 (2002).

76. Turchetti, M., Yang, Y., Bionta, M., Nardi, A., Daniel, L., Berggren, K. K. & Keathley, P. D. Electron emission regimes of planar nano vacuum emitters. *IEEE Transactions on Electron Devices* **69**, 3953–3959 (2022).
77. Laszczyk, K. & Krysztof, M. Electron beam source for the miniaturized electron microscope on-chip. *Vacuum* **189**, 110236 (2021).
78. Krysztof, M. Field-emission electron gun for a MEMS electron microscope. *Microsyst Nanoeng* **7**, 43 (2021).
79. Heo, S. H., Kim, H. J., Ha, J. M. & Cho, S. O. A vacuum-sealed miniature X-ray tube based on carbon nanotube field emitters. *Nanoscale Res Lett* **7**, 258 (2012).
80. Peralta, E. A., Soong, K., England, R. J., Colby, E. R., Wu, Z., Montazeri, B., McGuinness, C., McNeur, J., Leedle, K. J. & Walz, D. Demonstration of electron acceleration in a laser-driven dielectric microstructure. *Nature* **503**, 91–94 (2013).
81. Prat, E., Bettoni, S., Calvi, M., Dehler, M., Frei, F., Hommelhoff, P., Kozak, M., McNeur, J., Loch, C. O. & Reiche, S. Outline of a dielectric laser acceleration experiment at SwissFEL. *Nuclear Instruments and Methods in Physics Research Section A: Accelerators, Spectrometers, Detectors and Associated Equipment* **865**, 87–90 (2017).
82. O'Shea, B. D., Andonian, G., Barber, S. K., Fitzmorris, K. L., Hakimi, S., Harrison, J., Hoang, P. D., Hogan, M. J., Naranjo, B. & Williams, O. B. Observation of acceleration and deceleration in gigaelectron-volt-per-metre gradient dielectric wakefield accelerators. *Nature communications* **7**, 12763 (2016).